\documentclass[preprint,showpacs,preprintnumbers,amsmath,amssymb,prd]{revtex4}  
  
\usepackage{graphicx}
\usepackage{dcolumn}
\usepackage{bm}
  
\begin{document}  
  
\preprint{}  
  
\title{Cosmic microwave background constraints on a decaying
cosmological term related to the thermal evolution}
\author{Riou Nakamura $^1$}  
\author{Masa-aki Hashimoto $^1$}%
\author{Kiyotomo Ichiki $^2$}  
  
\affiliation{%
$^1$ Department of Physics, Graduate School of  
 Sciences,  Kyushu University,  
 4-2-1, Ropponmatsu, Chuo-ku, Fukuoka-city, Fukuoka 810-8560, Japan  
}%
  
\affiliation{  
$^2$ Research Center for the Early Universe,   
Graduate School of Science,   
The University of Tokyo,  
7-3-1, Hongo, Bunkyo-ku, Tokyo 113-0033, Japan  
}%
  
\date{\today}
\begin{abstract}  
 We constrain the thermal evolution of the universe with a decaying  
 cosmological term by using the method of the analysis for the Wilkinson
 Microwave Anisotropy Probe (WMAP) observation data.  The cosmological
 term is assumed  to be a function of the scale factor that increases
 toward the early universe, and the radiation energy density is lower
 compared to that in the model with the standard cosmological constant
 ($\Lambda$CDM). The decrease in the radiation density affects the
 thermal history of the universe; e.g. the photon decoupling occurs at
 higher-$z$ compared  to the case of the standard $\Lambda$CDM model.
As a consequence, a decaying cosmological term affects the cosmic 
 microwave background (CMB)  anisotropy. 
Thanks to the Markov-Chain Monte Carlo method, 
we compare the angular power spectrum in the decaying $\Lambda$CDM model
 with the CMB data, and we get severe constraints on parameters of the model.
\end{abstract}  
  
  
\maketitle  
  
\section{Introduction}  

Recent astronomical observations such as 
high redshift type Ia supernovae (SNIa) \cite{Perlmutter1998, Riess2007}, 
cosmic microwave background (CMB) anisotropy \cite{WMAP2006}, 
suggest the existence of dark energy strongly.
Although many researchers have investigated dark energy, 
its nature is still unknown.
In the proposed models or the equation of state of dark
energy \cite{Copeland2006}, the cosmological constant cannot be excluded \cite{Riess2007}.

If we assume that the dark energy is equivalent to a cosmological constant $\Lambda$,  
there rises again so called a \textit{cosmological constant problem}\cite{Weinberg1989}:
the present value of $\Lambda$ is extraordinarily small
compared with an inferred vacuum energy during the Planck time.
To solve this problem, it is natural to consider that 
$\Lambda$ decreases from a large value at the early epoch to the present value.   
Many functional forms of $\Lambda$ have been suggested : 
for instance 
decaying-$\Lambda$ has been introduced as a function of a
scalar field in the Brans-Dicke gravitational theory 
\cite{Nakamura2005}.
The evolution of the universe under various models of $\Lambda$ 
which are included in the energy-momentum tensor 
has been investigated analytically \cite{Overduin}.  
  
In addition, interacting $\Lambda$ with other kinds of 
energy has been also discussed.  
The vacuum energy of $\Lambda$ coupled with baryon could be
ruled out, because baryon-antibaryon created through vacuum decay
causes pair-annihilation. The produced high energy gamma ray flux
is contradicted with the observations \cite{Freese1986}.  
On the other hand, the vacuum energy decayed into the photon could affect
the cosmological evolution significantly. Assuming that the ratio of the
vacuum energy to the radiation is constant at the radiation dominated
era $(z>10^5_{})$, Freese et al. \cite{Freese1986} investigated the effects on
primordial nucleosynthesis and obtained a limit of vacuum to photon
energy ratio, which is less than 0.07. 
Furthermore, it is pointed out that observational constraints from the CMB
intensity put the limit on the ratio of the vacuum to the
radiation energy to be $\sim 10^{-3}_{}$  \cite{Overduin1993}. 

From the thermodynamical point of view, the temperature-redshift
relation is modified by including adiabatic photon creation due to
vacuum decay \cite{Lima1995, Lima1996}.
A phenomenological decaying-$\Lambda$ has been
found to affect the cosmological evolution after the recombination
\cite{Kimura, Kamikawa, Puy}. In models having $\Lambda$ terms as a
function of the scale factor,  the radiation and matter temperatures
would be significantly lower compared to the standard cold dark matter
model with a constant $\Lambda$ (S$\Lambda$CDM) \cite{Kimura}: the
molecular formation is occurred at earlier epoch 
by $\Delta z<10^3$ \cite{Kamikawa}. Furthermore, it is shown that in
some parameter regions, the radiation temperature could become
higher compared with the S$\Lambda$CDM model, which is found to be
consistent with the observational result of $z<4$ \cite{Puy}.  
  
Related to  the recent observations, the first star formation that
occurred at the end of the {\it dark age} has been investigated
progressively \cite{Bromm2003}.   
In decaying $\Lambda$ models, using a cooling diagram,
the first star formation was estimated to occur at an earlier epoch by
$\Delta z\sim 20$ with its mass $\sim10^6_{}M^{}_{\odot}$ \cite{Kamikawa}.  
In the meantime, from an observational approach, the CMB polarization  
observed by the WMAP satellite predicts  via measured reionization redshift
with use of the S$\Lambda$CDM model that a first object was formed
around $z =10$ \cite{WMAP2006, Page06}.
Since the CMB anisotropies give severe constraints on parameter regions
concerning the cosmological evolution,
we can also estimate the era of the first star formation
in the decaying-$\Lambda$ cosmology using the CMB anisotropy.  

In the present paper, we constrain the parameter regions that determine
the thermal history of the universe with a $\Lambda$ decaying into
the photon (hereafter we call it D$\Lambda$CDM ). In Sec. \ref{sec:model}, 
we describe briefly the thermal evolution and clarify the effects on
photon decoupling in the D$\Lambda$CDM model. In Sec. \ref{sec:cmba} we
examine the consistency between D$\Lambda$CDM and the CMB anisotropy
data of Winlinson Microwave Anisotropy Probe (WMAP) using the
Markov-Chain Monte Carlo (MCMC) method.
Summary and discussion are given in Sec. \ref{sec:rad}.  
  
\section{Thermal evolution with a decaying cosmological term}  
{\label{sec:model}}  

Using the Friedmann-Robertson-Walker metric,   
the Einstein equation and/or the energy-momentum conservation law   
are written as follows:  
\begin{eqnarray}  
\left( \frac{\dot{a}}{a} \right)^2_{} &=&  
 \frac{8\pi G}{3}\bar{\rho}a^2_{}-K, 
 \label{eq:friedeq} \\  
 \dot{\bar{\rho}} &=&-3\frac{\dot{a}}{a}\left( \bar{\rho}+\bar{p} \right),  
\label{eq:rhodot}  
\end{eqnarray}  
where $a, K$ and $G$ are the cosmic scale factor, the curvature and the
gravitational constant, respectively.
We note that bars such as $\bar{\rho}$ and $\bar{p}$ indicate the
average values during the cosmological evolution.
We choose the unit such that the velocity of light $c = 1$.
Note that {\it dot}s in Eqs. (\ref{eq:friedeq}) and (\ref{eq:rhodot})
indicate the derivative concerning a conformal time $\tau$. The total energy density $\bar{\rho}$ and the  
pressure $\bar{p}$ are written as  
\begin{equation}  
 \bar{\rho} = \bar{\rho}^{}_m+\bar{\rho}^{}_\gamma+\bar{\rho}^{}_{\nu}+\bar{\rho}^{}_{\Lambda}, \qquad  
  \bar{p} = \bar{p}^{}_{\gamma}+\bar{p}^{}_{\nu}+\bar{p}^{}_{\Lambda},  
  \label{eq:rhop}  
\end{equation}  
where the subscripts $m, \gamma$, $\nu$, and $\Lambda$ indicate the nonrelativistic matter  
(baryon plus cold dark matter), photon, neutrino, and a cosmological term, respectively.  
The equation of states $\bar{p}/\bar{\rho}$   
for individual components are written as, 
\begin{equation}  
\bar{p}/\bar{\rho}=  
\begin{cases}  
 1/3 & \text{relativistic particles},     \\  
 0   & \text{non-relativistic particles}, \\  
 -1  & \text{cosmological term }.  
\end{cases}  
\label{eq:eoslist}  
\end{equation}  
Here the energy densities of matter and neutrino   
vary as   
$\bar{\rho}^{}_m=\bar{\rho}^{}_{m0}a^{-3}_{}$ and $\bar{\rho}^{}_{\nu}=\bar{\rho}^{}_{\nu0}a^{-4}$,   
where the subscript $0$ means the present value.  
  
From Eqs.~(\ref{eq:rhodot}), (\ref{eq:rhop}) and (\ref{eq:eoslist}),   
we get the evolution equation of the photon energy density after   
the epoch of electron-positron pair-annihilation :  
\begin{equation}  
  \frac{d\Omega^{}_{\gamma}}{da}+4\frac{\Omega^{}_{\gamma}}{a}  
	=-\frac{d\Omega^{}_{\Lambda}}{da},  
 \label{eq:emcona}  
\end{equation}  
with the density parameter $\Omega^{}_{i}$ 
\[  
\Omega^{}_{i} = \frac{\bar{\rho}^{}_{i}}{\rho^{}_{\mathrm{crit}}} ,   
\quad \rho^{}_{\mathrm{crit}} = \frac{3 H_{0}^{2}}{8 \pi G} ,  
\]  
where $H^{}_0$ is the Hubble constant in units of km/sec/Mpc.  
  
In the D$\Lambda$CDM model, the evolution of the photon is affected by the time-dependent
cosmological term. In this work, we assume a functional form of
$\Lambda$ as follows \cite{Matyjasek1994,Kimura, Puy,Kamikawa}:   
\begin{equation}  
 \Omega^{}_{\Lambda}=\Omega^{}_{\Lambda 1}+\Omega^{}_{\Lambda 2}a^{-m}, \label{eq:lambda}  
\end{equation}  
where $\Omega^{}_{\Lambda1},\Omega^{}_{\Lambda2}$ and $m$ are {} constants.  
Note that the present value of $\Omega^{}_{\Lambda}$ is expressed by   
$\Omega^{}_{\Lambda0}=\Omega^{}_{\Lambda 1}+\Omega^{}_{\Lambda 2}$.  
Formalism of this paper is based on previous studies in Refs.\cite{Kimura,Kamikawa}.  
  
Integration of  Eq.~(\ref{eq:emcona}) with (\ref{eq:lambda}) and 
Stefan-Boltzmann's law,  $\bar{\rho}^{}_{\gamma}\propto T^4_{\gamma}$, 
leads to the following photon temperature as a function of the scale
factor \cite{Puy},
\begin{equation}  
  T^{}_{\gamma} = \frac{T^{}_{\gamma 0}}{a}\times  
\begin{cases}  
{\left[ 1+\dfrac{\alpha}{\Omega^{}_{\gamma 0}}\left( a^{4-m}_{}-1 \right)  
\right]}^{1/4}_{} & (m\ne 4), \\  
\left( 1+4\dfrac{\Omega^{}_{\Lambda 2}}{\Omega^{}_{\gamma 0}}  
\ln{a}\right)^{1/4}_{} & (m=4)   
\end{cases}  
\label{eq:Tgamma}  
\end{equation}  
where $T^{}_{\gamma}$ is the present photon temperature and 
 $\alpha\equiv{m\Omega_{\Lambda2}}/{(4-m)}$.
The present photon energy density 
$\Omega^{}_{\gamma 0}=2.471\times 10^{-5}_{}h^{-2}_{}( T^{}_{\gamma 0}/2.725 {\rm ~K})^4$  
with the normalized Hubble constant $h$ ($H^{}_{0} = 100 \, h$ km/sec/Mpc).
Second terms in Eq. (\ref{eq:Tgamma}) are characteristic ones in the D$\Lambda$CDM
model; the evolution of the photon is modified by those terms.
 
\begin{figure}[t]  
\includegraphics[width=1.0\linewidth]{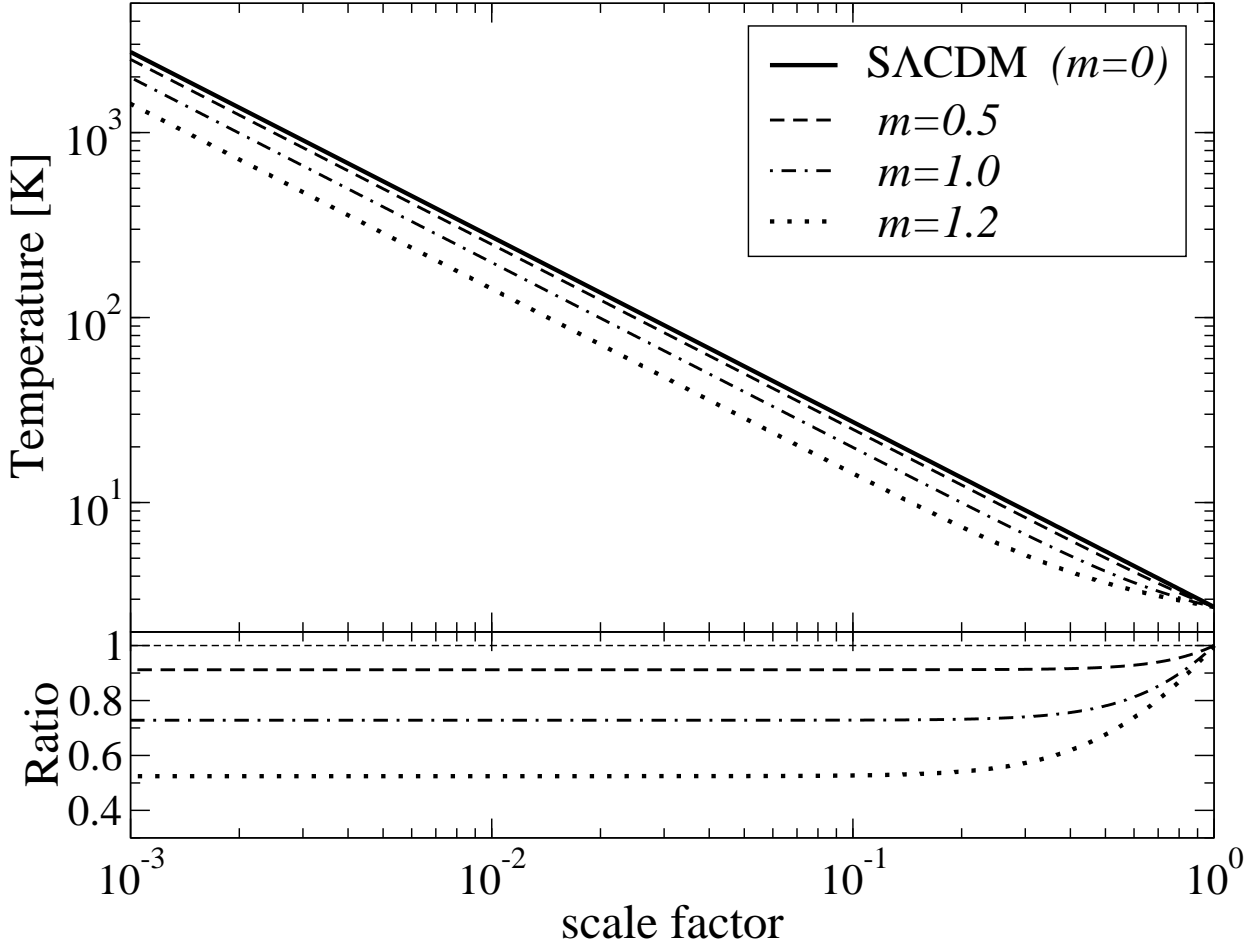}
\caption{\label{fig:temp_dlcdm}   
Upper panel: the evolution of the photon  
 temperature in D$\Lambda$CDM ($m=0-1.2$) with $\Omega^{}_{\Lambda2}=10^{-4}$ after  
 the hydrogen  recombination era.  Lower panel: the ratios of $T_\gamma$ to  
 that of S$\Lambda$CDM.}  
\end{figure}

Figure \ref{fig:temp_dlcdm} illustrates the evolution of the photon  
temperature in the D$\Lambda$CDM model. It can be seen that $T_{\gamma}$ in
D$\Lambda$CDM is lower compared to that in S$\Lambda$CDM.   
For $0<m<4$, the photon evolves as $T^{}_{\gamma}\propto a^{-1}_{}$ at  
the early epoch and the slope of $T^{}_{\gamma}$ against $a$ decreases
due to the contribution of $a^{4-m}_{}$ near the present epoch.  
For $m>4$, the opposite results are obtained.  
Therefore, decaying-$\Lambda$ affects the thermal history at around the
present epoch of $z<10^3$ and particularly after the hydrogen
recombination era.  

If $\Omega^{}_{\Lambda 2}$ and/or $m$ is very large, 
the solution with present $T_{\gamma0}$ fixed indicates that
the total energy density becomes negative for some epoch of $z>0$. As a
consequence, these parameters were constrained as  
$m\Omega_{\Lambda2}\leq 10^{-3}$ \cite{Kimura}. 
In our analysis, 
the photon temperature for large $m$ and/or $\Omega^{}_{\Lambda2}$ 
becomes also negative at some epoch of $a<1$. By excluding this kind of 
solution, we obtain the upper limits on both $\Omega^{}_{\Lambda 2}$ and $m$ 
from Eq.~(\ref{eq:Tgamma}): 
\begin{equation} 
\alpha < \Omega^{}_{\gamma 0} \quad ( m<4 ). 
\label{eq:upperlimits} 
\end{equation} 
In the case of $m\geq4$, we assume $T^{}_{\gamma}>0$ until the primordial 
nucleosynthesis epoch, $a=10^{-10}_{}$, and obtain the limits 
\begin{align} 
 \Omega^{}_{\gamma0} &\gtrsim 92 \Omega^{}_{\Lambda2}, \quad (m=4) 
 \label{eq:upperlimits_eq4}\\ 
\Omega^{}_{\gamma0} &> -10^{10(m-4)}_{}\alpha 
 \quad (m>4) 
 \label{eq:upperlimits_over4}
\end{align}
On the other hand, for $\Omega^{}_{\Lambda 2}<0$ or $m<0$, we find that
$T^{}_{\gamma}$ becomes negative at some time of $a>1$. Therefore we
impose the two conditions of $\Omega^{}_{\Lambda 2}\geq 0$ and $m\geq0$. 

In the D$\Lambda$CDM model, the cosmological term decreases from the early  
time to the present, because the second term in Eq.~(\ref{eq:lambda})  
becomes larger than the first one at the early epoch of 
$z>0$. 
Since the first term in  
Eq.~(\ref{eq:lambda}) dominates near the present epoch, the $\Lambda$  
term is nearly constant for low-$z$. 
Although cosmological models with the $\Lambda$ term have been tightly
constrained from the luminosity-redshift relation of SNIa,  
effects on the expansion rate  are negligible in the D$\Lambda$CDM model.

In the S$\Lambda$CDM model, the ratio of the photon to the neutrino
temperature is $T^{}_{\gamma}/T^{}_{\nu}=(11/4)^{1/3}$ after
electron-positron annihilation, because $T^{}_{\gamma}$ and $T^{}_{\nu}$
evolve as $\propto a^{-1}_{}$. In the D$\Lambda$CDM model, since a
decaying-$\Lambda$ alters the evolution of the photon, the ratio of
the photon-to-neutrino depends on time. Nonetheless if we set
$T^{}_{\gamma}/T^{}_{\nu}=(11/4)^{1/3}$ at $a=10^{-10}_{}$ in the
D$\Lambda$CDM model, the present neutrino temperature is lower than that
in S$\Lambda$CDM as seen in the second column of Table \ref{tab:temp_dec}. 
Although recent observational studies put constraints on properties of
the cosmic neutrino background such as the neutrino species or masses
(e.g. \cite{WMAP2006, Fukugita2006}), there is no observation about its
temperature (or the energy density). Therefore, D$\Lambda$CDM seems to
have no problems for the lower neutrino temperature.  

\begin{figure}[p]
\begin{center}
 \includegraphics[width=.9\linewidth,keepaspectratio]{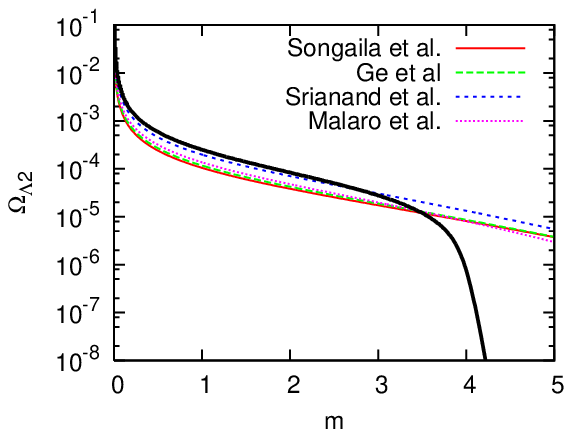}
 \includegraphics[width=.9\linewidth,keepaspectratio]{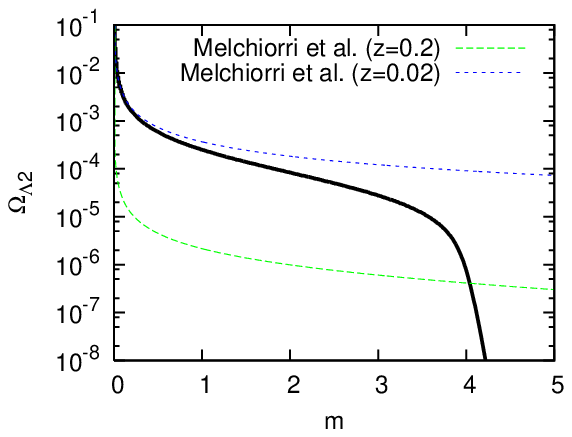}
\end{center}
\caption{\label{fig:m_oml2_temp} Constraints on the $m-\Omega^{}_{\Lambda2}$ plane
 from observational temperatures. The black-solid line shows the upper
 limits of  parameters by Eqs. (\ref{eq:upperlimits}), (\ref{eq:upperlimits_eq4}), and
 (\ref{eq:upperlimits_over4}), and the other lines indicate 
 the upper limits obtained from the same analysis in Ref.\cite{Puy} . 
Upper panel: constraints from the temperature at $z>1$. 
Lower panel: constraints from the temperature at $z<1$. }
\end{figure}

CMB temperature at $z=0$ was measured accurately by the Far Infrared
Absolute Spectrophotometer of Cosmic Background Explorer: $T=2.725\pm0.002$ K at
2$\sigma$ C.L. \cite{Mather1999}. 
On the other hand, CMB temperature observation at $z>0$ are reported in
literature \cite{Songaila1994, Battistelli2003}.
Consistency of the temperature evolution in the D$\Lambda$CDM model with
these observational results has been discussed \cite{Puy}.
When $m$ and/or $\Omega^{}_{\Lambda2}$ take a large value, $T^{}_{\gamma}$ in
the D$\Lambda$CDM model is not consistent with the temperature observation.
Puy \cite{Puy} has put constraints only on the  $m-\Omega^{}_{\Lambda2}$
plane from the temperature observation of $z<1$ \cite{Battistelli2003} : 
\begin{equation}
 |m| \le 1, |\Omega^{}_{\Lambda 2}|\le 10^{-4}_{}.
\label{eq:by_temp_limits}
\end{equation}
These limits are obtained by comparing the observational temperature
included $1\sigma$ error with the temperature given in Eq. (\ref{eq:Tgamma}). 
Thus, we perform the extended analysis to constrain the parameters using
the available observations as precisely as possible.
Figure \ref{fig:m_oml2_temp} shows constraints on 
the $m-\Omega^{}_{\Lambda2}$ plane from the observational temperature 
using the same analysis in Ref. \cite{Puy} under the theoretical request
of $T_{\gamma}>0$.
Constraints from $T_{\gamma}$ at $z>1$ are similar for $m>4$ as shown in
the upper panel of Fig.\ref{fig:m_oml2_temp} and those from $T_{\gamma}$
at $z<1$ have large uncertainty as shown in the lower panel of
Fig.\ref{fig:m_oml2_temp}. These limits of $m$ or
$\Omega^{}_{\Lambda2}$ obtained from the observed temperatures are
consistent with the excluded region by Eqs.(\ref{eq:upperlimits}),
(\ref{eq:upperlimits_eq4}) and (\ref{eq:upperlimits_over4}).
Since obtained constraints are rather rough, we put further severe
limits using CMB anisotropy as shown in Sec.\ref{sec:cmba}.

\begin{figure}[t]  
 \includegraphics[width=1.0\linewidth]{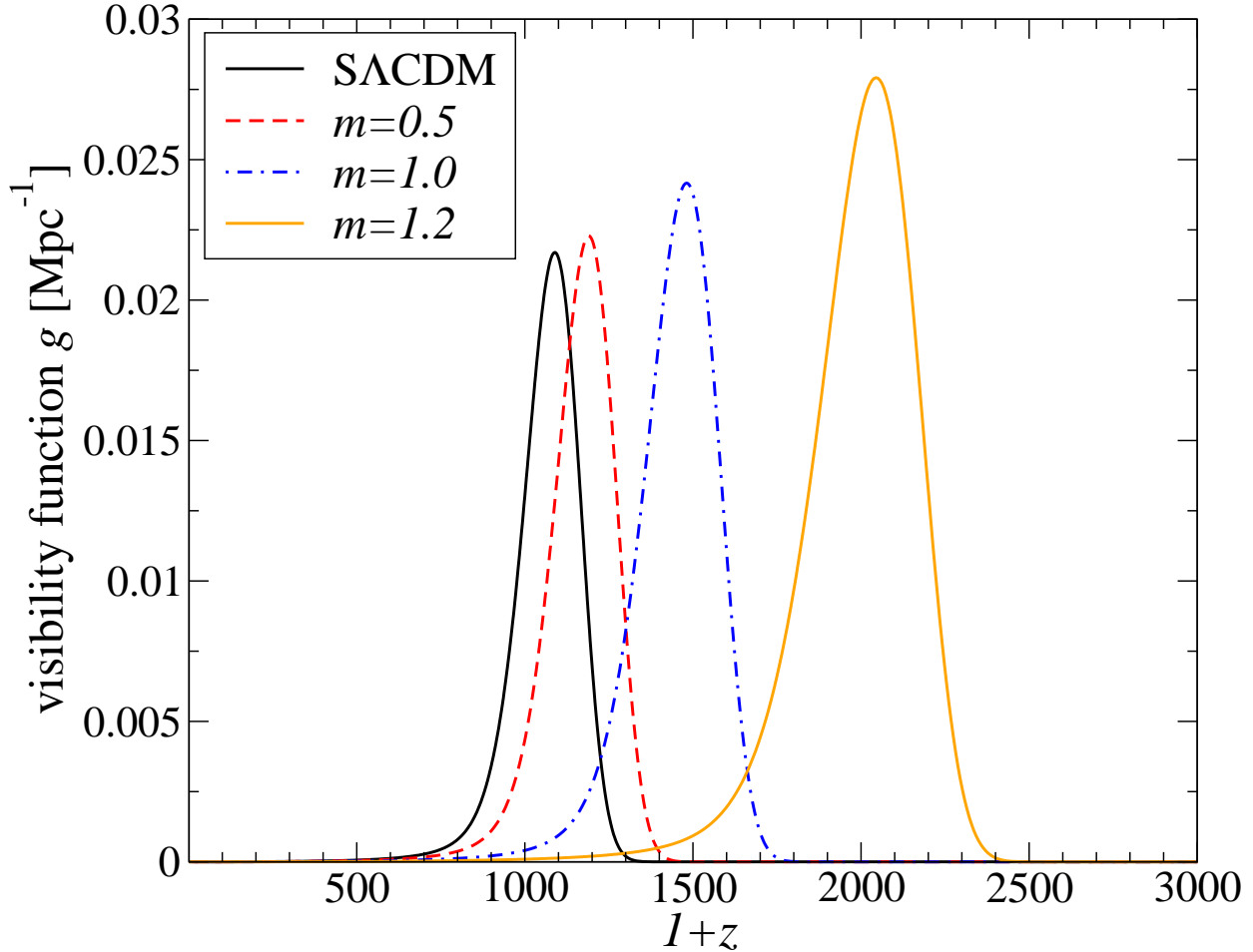}  
  \caption{\label{fig:visfunc_dlcdm} Visibility function  
 vs the redshift for a fixed parameter
 $\Omega^{}_{\Lambda2}=10^{-4}_{}$. Peaks of $g$ show the epoch of
 photon decoupling.  
 As the values of the parameter $m$ increase, the photon decoupling
 occurs at an earlier epoch.}  
\end{figure}  

Now we estimate the epoch of photon decoupling by   
calculating the visibility function $g(t)$ which has a peak at the epoch  
of the last-scattering surface:
\[  
  g(t)=-\dot{\tau}e^{-\tau}, \qquad  
  \tau=\sigma^{}_{T}\int{n^{}_{e}dt},  
\]  
where $\sigma_{T}$ is the Thomson-scattering cross-section   
and $n_e$ is the number density of the free electrons  
that depends on the recombination history of the universe.
  
Figure \ref{fig:visfunc_dlcdm} illustrates the visibility function  
as a function of redshift in the S$\Lambda$CDM and D$\Lambda$CDM models. The
epoch of photon decoupling shifts to higher-$z$ as shown in the third
column of Table \ref{tab:temp_dec}. 
When we take the upper limits obtained from Eq.~(\ref{eq:upperlimits}),     
 $(\Omega_{\Lambda2},m)=(10^{-4},1.2)$,     
the photon decoupling occurs at $z_{dec}=2040$ that is earlier     
by $\Delta z_{dec}\sim950$ compared to the case of the S$\Lambda$CDM.
In addition, we find that  photon temperature at the last-scattering
surface is about 0.1\% lower than that in the S$\Lambda$CDM model, as
shown in the fourth column of Table \ref{tab:temp_dec}.
These effects should be further constrained by the cosmological
observations such as CMB anisotropy as shown in the following sections.    

\begin{table} 
\caption{Present neutrino temperature ($z=0$), redshift, and  photon temperature at
the last-scattering surface ($z_{dec}$) in the decaying-$\Lambda$ model
 with $\Omega_{\Lambda2}=10^{-4}_{}$.}    
\label{tab:temp_dec}     
\begin{ruledtabular}    
\begin{tabular}[t]{ccccc}    
Parameter & $T_\nu$ [K] ($z=0$)  & $z^{}_{dec}$ & $T^{}_{\gamma}$ [K] ($z_{dec}$) &     
 \\    
\hline    
  $m=0.0$ & $1.945$ & $1087$ &  $2965$  
\\    
  $m=0.5$ & $1.774$ & $1188$ &  $2957$  
\\    
  $m=1.0$ & $1.416$ & $1480$ &  $2939$  
\\    
  $m=1.2$ & $1.022$ & $2043$ &  $2921$  
\\    
\end{tabular}    
\end{ruledtabular}    
\end{table}    
    
\section{\label{sec:cmba} Effects on CMB anisotropy and constraints by
 Markov-Chain Monte Carlo analysis}    
    
Before discussing effects of a decaying-$\Lambda$ on the CMB power    
spectrum, let us formulate the Boltzmann equation for the photon in    
the D$\Lambda$CDM model    
based on the cosmological perturbation theory.     
 The line element in the synchronous gauge with the flat space is written as     
 \begin{equation}    
   ds^{2}=a(\tau)\left[    
  -d\tau^2+\left( \delta_{ij}+h_{ij}\right)dx^idx^j    
 \right]    ,
 \label{eq:lineelement}    
 \end{equation}    
where $h^{}_{ij}$ is the metric perturbation.    
 We introduce two fields $h(\bm{k},\tau)$ and $\eta(\bm{k},\tau)$    
 in the Fourier $k$-space and write the scalar mode of  $h^{}_{ij}$ as the Fourier    
 integral: 
 \[    
  h_{ij}(\bm{x},\tau)=\int{d^3k e^{i\bm{k}\cdot\bm{x}}_{}    
 \left[\hat{\bm{k}}^{}_i \hat{\bm{k}}^{}_j h(\bm{k},\tau)+\left(    
 \hat{\bm{k}}^{}_i \hat{\bm{k}}^{}_j -\frac{\delta^{}_{ij}}{3} \right)6\eta(\bm{k},\tau) \right]    
  }    ,
 \]    
where $\bm{k}^{}_{j}=k \hat{\bm{k}}^{}_j$ with the unit vector $\hat{\bm{k}}^{}_j$ .     
Components of the energy-momentum tensor with perturbed parts are given by     
\begin{equation}
\begin{split}
  T^0_0 &= -\left(\bar\rho+\delta\rho\right), \\     
  T^0_j &= -T^j_0=\left(\bar\rho+\bar{p}\right)v_j,  \\     
  T^i_j &= \left( \bar{p}+\delta p \right)\delta^i_j+\Sigma^i_j,    
\end{split}
 \label{eq:energytensor}    
\end{equation}
 where $\delta\rho, \delta p, v_i$ and $\Sigma^i_j$ are perturbed parts
  of the energy density, pressure,   
 velocity of fluids, and  anisotropic stress, respectively.    

In the D$\Lambda$CDM model, the energy density of the photon in the
background part is obtained from Eq. (\ref{eq:emcona}).    
 The perturbed parts of the equation of the energy-momentum conservation
 that correspond to the first order perturbation reduce to the following
 equations, 
\begin{eqnarray}    
\dot{\delta}_{\gamma} & = &    
 -\frac{4}{3}\theta^{}_{\gamma}-\frac{2}{3}\dot{h}+\frac{\dot{\rho}^{}_{\Lambda}}{\bar{\rho}^{}_\gamma}\delta^{}_\gamma    
 \label{eq:deltag_dlcdm} \\    
\dot{\theta}^{}_{\gamma} & = &    
\frac{1}{4}k^2\delta^{}_{\gamma}-k^2\sigma_{\gamma}+\frac{\dot{\rho}^{}_{\Lambda}}{\bar{\rho}^{}_\gamma}\theta^{}_\gamma ,    
\label{eq:thetag_dlcdm}    
\end{eqnarray}    
where  $\delta_{\gamma}=\delta\rho_{\gamma}/\bar{\rho_{\gamma}}$ and
$\theta_{\gamma}$ is the divergence of    
the fluid velocity, $\theta_{\gamma} \equiv ik^jv_j$.
$\sigma^{}_{\gamma}$ is defined by  
\[    
 \left( \bar{\rho_{\gamma}}+\bar{p_{\gamma}}\right)\sigma_{\gamma}\equiv    
-\left(\hat{\bm k}_i\cdot\hat{\bm k}_j-\frac{1}{3}\delta_{ij} \right)\Sigma^i_j.    
\]    
Equations (\ref{eq:deltag_dlcdm}) and (\ref{eq:thetag_dlcdm}) are the continuity    
and Euler equations, respectively. Note that we take into account the
interaction terms proportional to $\dot{\rho}_{\Lambda}$ in these
equations
and the effects will be small since the ratio
$\rho^{}_{\Lambda}/\rho{}_{\gamma }$ at $z=10^3_{}$ is less than $10^{-3}_{}$.

Moreover, to construct the perturbed evolution equation of the photon, we need
the contribution of the higher multipole moments. 
The Boltzmann equation for a relativistic particle in $k$-space is written
as follows \cite{Ma1995} :
\begin{equation}    
 f^{}_{0}\frac{\partial\Psi}{\partial\tau}+\Psi\frac{\partial f^{}_{0}}{\partial\tau}    
  +ik\mu f^{}_0\Psi+\frac{d \ln{f^{}_0}}{d \ln{q}}    
  \left( \dot{\eta}+\frac{\dot{h}+6\dot{\eta}}{2}\mu^2_{}\right)=    
  \left( \frac{\partial f^{}_{}}{\partial\tau}\right)^{}_{col},     
  \label{eq:BoltzmannEq}    
\end{equation}    
where $\mu=\hat{k}\cdot \hat{n}$, and $q^{}_i=qn^{}_{i}$ is the 3-dimensional momentum.    
The right-hand side of Eq. (\ref{eq:BoltzmannEq}) is the collision term.    
The distribution function expressed by the convolution of 
zeroth-order and the perturbed part is written as
\[    
 f^{}_{}(x^i,q,n_j,\tau)=f^{}_{0}(q,\tau)(1+\Psi^{}_{}(x^i,q,n_j,\tau)).    
\]    
In the S$\Lambda$CDM model, the zeroth-order distribution function,
$f^{}_{0}$, of the photon is described as
$f^{}_{\gamma0}(q)=1/{(h^{}_p\exp{(q/k_BT_{\gamma})}- 1)}$ ,
where $h^{}_p$ and $k^{}_B$ are the Planck and Boltzmann constants,
respectively. 
However, in models with a created photon by decaying vacuum energy, 
the spectral distribution of CMB is a function of both temperature and
the comoving number of photon,
where $f_0$ takes a generalized Planckian form \cite{Lima1995}.
Therefore, we cannot drop the second term in the left-hand side in
Eq. (\ref{eq:BoltzmannEq}),
which is the time derivative of $f_0$.

To obtain the Boltzmann equation for photon, we expand the angular    
dependent part of the perturbation in a series of Legendre polynomials $P^{}_l(\hat{k}\cdot\hat{n})$ as
follows:
\[    
 F^{}_{\gamma}(\bm{k},\hat{n},\tau )\equiv    
\frac{\int{q^2_{}dqf^{}_{0}(q,\tau)\Psi}}{\int{q^2_{}dqf^{}_{0}(q,\tau)}}    
=\sum_{l=0}^{\infty}{(-i)^l_{}\left(2l+1\right)F^{}_{\gamma l}(\bm{k},\tau)P^{}_{l}(\mu)}.    
\]
We integrate Eq.(\ref{eq:BoltzmannEq}) multiplied by $q^3_{}dqf^{}_0$ 
over the whole $p$ space and divide it by $\int{q^3_{}dqf^{}_{0}}$.
Then we obtain the following Boltzmann equation for the CMB photon in
$k$-space:
\begin{equation}    
 \dot{F}^{}_{\gamma}+ik\mu F^{}_{\gamma}+    
\frac{4}{3}\left( \dot{h}+6\dot{\eta}\right) P^{}_2(\mu)+\frac{2}{3}\dot{h}    
-\frac{\dot{\rho}^{}_{\Lambda}}{\bar{\rho}^{}_{\gamma}}F^{}_{\gamma}    
=\left( \frac{\partial F^{}_{\gamma}}{\partial\tau}\right)^{}_{col} .    
\label{eq:Boltzmann2}    
\end{equation}    
The last term of the left-hand side in Eq. (\ref{eq:Boltzmann2})     
corresponds to a new one that appeared in the D$\Lambda$CDM model.    
The collision term in the right-hand side is described as Thomson scattering \cite{Ma1995}:    
\begin{equation}    
 \left( \frac{\partial F^{}_{\gamma}}{\partial\tau} \right)^{}_{col}    
 = an^{}_{e}x^{}_{e}\sigma^{}_{T}\left[     
-F^{}_{\gamma}+F^{}_{\gamma0}+4\hat{n}\cdot\bm{v}^{}_{e}    
-\frac{1}{2}\left( F^{}_{\gamma}+G^{}_{\gamma0}+G^{}_{\gamma2}\right)P^{}_{2}(\mu)    
\right],  
\end{equation}
where $G^{}_{\gamma l}$ is the difference of the two linear polarization components.
   
Substituting the Legendre expansion for $F^{}_{\gamma}$, and 
using the orthonormality of the Legendre polynomial with the  recursion relation    
$(l+1)P^{}_{l+1}(\mu)=(2l+1)\mu P^{}_l(\mu)-lP^{}_{l-1}(\mu)$ , 
we get the Boltzmann equations for the photon in the D$\Lambda$CDM model as follows:
\begin{eqnarray}    
\dot{\theta}^{}_{\gamma}&=&    
\frac{1}{4}k^2\delta^{}_{\gamma}-k^2\sigma^{}_{\gamma}    
-an^{}_{e}x^{}_{e}\sigma^{}_{T}\left( \theta^{}_{\gamma}-\theta^{}_{b}\right)    
+\frac{\dot{\rho}_{\Lambda}}{\bar{\rho}^{}_\gamma}\theta^{}_\gamma  ,   
\label{eq:thetag_dlcdm2} \\    
\dot{\sigma}^{}_{\gamma}&=&\frac{4}{15}\theta^{}_{\gamma}-\frac{3}{10}kF^{}_{\gamma3}    
 +\frac{2}{15}\left( \dot{h}+6\dot{\eta}\right)    
 -\frac{9}{10}an^{}_{e}x^{}_{e}\sigma^{}_{T}\sigma^{}_{\gamma}    
 +\frac{1}{20}an^{}_{e}x^{}_{e}\sigma^{}_{T}\left( G^{}_{\gamma0}+G^{}_{\gamma2}\right)    
+\frac{\dot{\rho}^{}_{\Lambda}}{\bar{\rho}^{}_\gamma}\sigma^{}_\gamma    ,
\label{eq:sigmag_dlcdm}\\    
\dot{F}^{}_{\gamma l}&=&\frac{k}{2l+1}\left[     
lF^{}_{\gamma,l-1}-\left(l+1\right)F^{}_{\gamma, l+1}    
\right]    
 -an^{}_{e}x^{}_{e}\sigma^{}_{T}F^{}_{\gamma l}      
+\frac{\dot{\rho}^{}_{\Lambda}}{\bar{\rho}^{}_{\gamma}}F^{}_{\gamma l}    
~~(l\geq3)     , 
\label{eq:Fgl_dlcdm}    
\end{eqnarray}    
where    
\[    
\delta^{}_{\gamma}= F^{}_{\gamma0}    
,~\theta^{}_{\gamma}=\frac{3}{4}kF^{}_{\gamma1}    
,~\sigma^{}_{\gamma}=\frac{F^{}_{\gamma2}}{2}.    
\]    
    
\begin{figure}[htb]    
\begin{center}    
 \includegraphics[width=1.\linewidth,keepaspectratio]{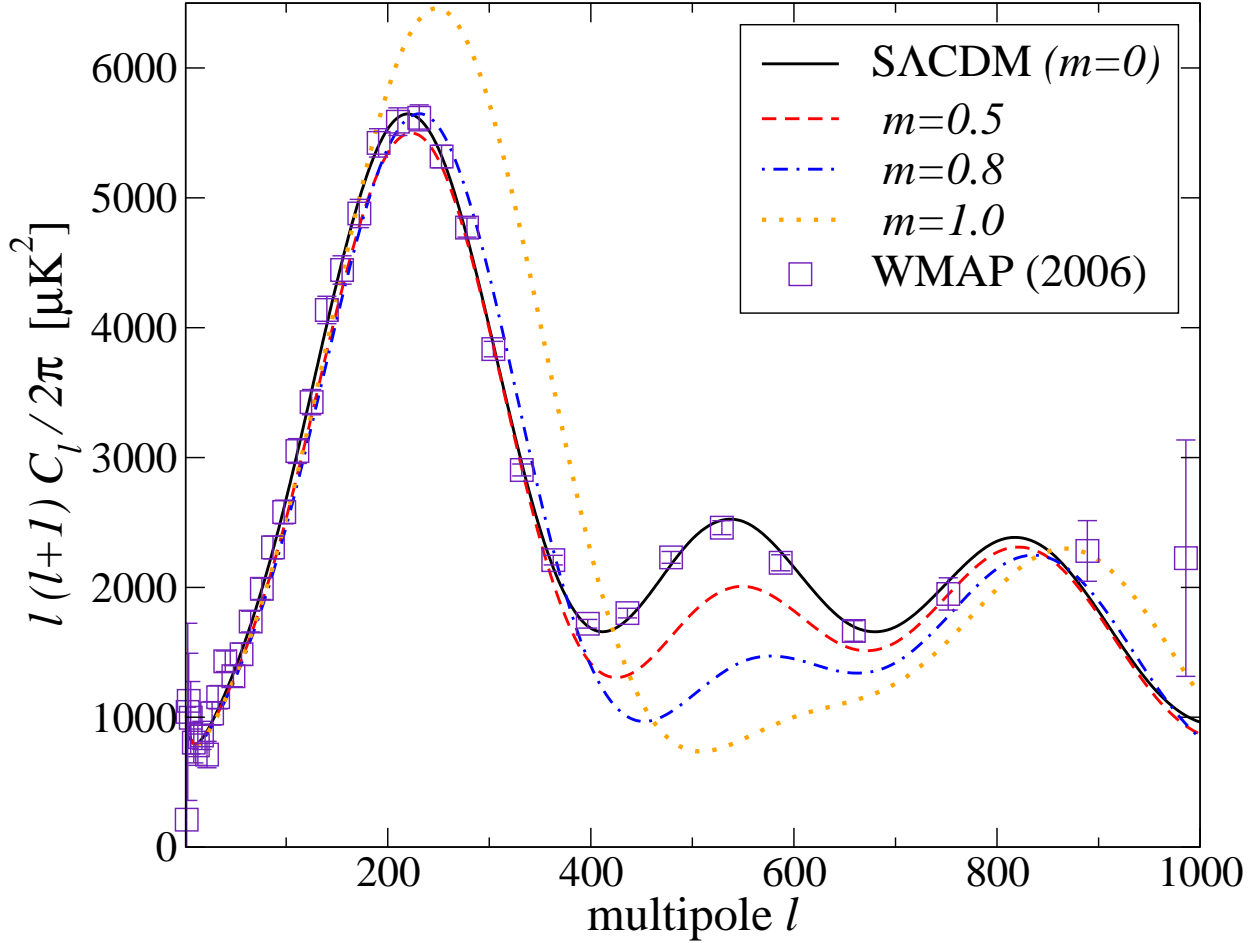}    
 \caption{\label{fig:cl_dlcdm} Comparison of the angular power spectrum
 in the decaying $\Lambda$ model with the WMAP observation data. The
 solid line is the result of S$\Lambda$CDM. The dashed,  dot-dashed, and
 dotted lines are those of D$\Lambda$CDM with
 $(\Omega^{}_{\Lambda 2},m)=(10^{-4}, 0.5)$, $(10^{-4}, 0.8)$, and
 $(10^{-4}, 1.0)$, respectively.}    
\end{center}    
\end{figure}

We calculate the CMB power spectrum by modifying the CAMB code \cite{Lewis1999}     
 based on the CMBFAST code \cite{Seljak1996}, where we include 
the modified Boltzmann equation of
 Eqs. (\ref{eq:deltag_dlcdm}), (\ref{eq:thetag_dlcdm2})--(\ref{eq:Fgl_dlcdm}).
Figure \ref{fig:cl_dlcdm} shows the effects of $m$ on the angular power 
spectrum with the following cosmological parameters: the baryon density
 parameter $\Omega^{}_bh^2_{}=0.0223$, 
the cold dark matter (CDM) density parameter $\Omega^{}_{CDM}h^2_{}=0.104$,     
$K=0$, $h=0.73$, and the reionization is neglected.    
We find that a decaying-$\Lambda$ modifies the CMB power spectrum as
follows: 
if $m$ and/or $\Omega^{}_{\Lambda2}$ is small, the amplitude of the power    
spectrum decreases. If we take larger values of $m$ and/or $\Omega^{}_{\Lambda2}$,     
the first and third peaks of the power spectrum increase    
due to the large baryon density relative to    
the photon energy density.     
Furthermore, the CMB power spectrum shifts toward    
higher-$l$, because the photon last-scattering occurs at an earlier    
epoch as seen in Fig.~\ref{fig:visfunc_dlcdm} and Table~\ref{tab:temp_dec}.  
We have found that the new term in the D$\Lambda$CDM model increases the angular
power spectrum at $l>20$  by about $10^{-3}_{}$ percents.
Therefore, the observational constraints are the same
even if these terms are not included.

    
The CMB angular power spectrum is rather sensitive to other cosmological parameters.    
For instance, baryon and CDM densities affect the amplitude of CMB anisotropy.    
Therefore, we need to carry out the MCMC approach \cite{Lewis2002}    
to constrain the possible parameters:    
$\Omega^{}_bh^2$, $\Omega^{}_{CDM}h^2$,  $h$,    
the reionization redshift $z^{}_{re}$,     
the scalar spectral index $n^{}_s$,  
the amplitude of density fluctuation $A^{}_s$,     
and two parameters in D$\Lambda$CDM ($\Omega^{}_{\Lambda2}$ and $m$).    
We note that we do not assume flat universe.

To start the MCMC calculations,    
we assume the priors on the cosmological parameters as follows:    
$0.5\leq n^{}_s\le 1.5$,  
$\Omega^{}_bh^2_{}= 0.022\pm 0.0022$ at $1\sigma$ C.L. (BBN prior),
$0.01\le \Omega^{}_{CDM}h^2_{}\le0.99$,
$-0.3\leq\Omega^{}_{K0} (\equiv-K/H^2_0)\leq 0.3$,     
$0\le\Omega^{}_{\Lambda1}\leq 1.0$     
, and $10$ Gyr $< t_0 < 20$ Gyr (age of the universe).    
    
\begin{figure}[t]
\begin{center}
 \includegraphics[width=1.0\linewidth]{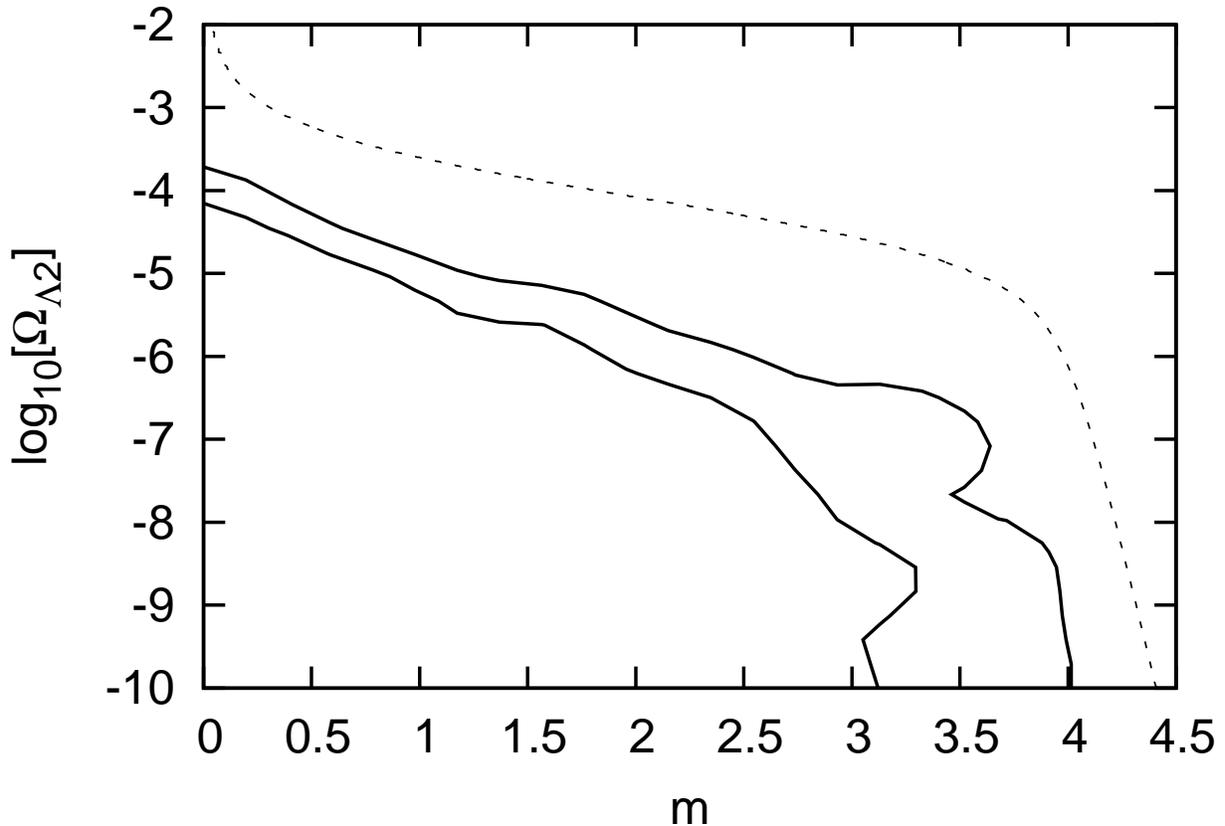}
\caption{\label{fig:cntr_dlcdm} 
Contours of the $m-\Omega^{}_{\Lambda2}$ plane from CMB. Soqlid lines
 indicate 68.3\% and 95.4\% confidence levels.
The dotted-line is the upper-limit from  Eqs. (\ref{eq:upperlimits}),
 (\ref{eq:upperlimits_eq4}), and (\ref{eq:upperlimits_over4}) with $h=0.73$.
} 
\end{center}
\end{figure}

\begin{figure*}[t]    
\begin{minipage}{0.49\linewidth}    
 \includegraphics[width=1.0\linewidth]{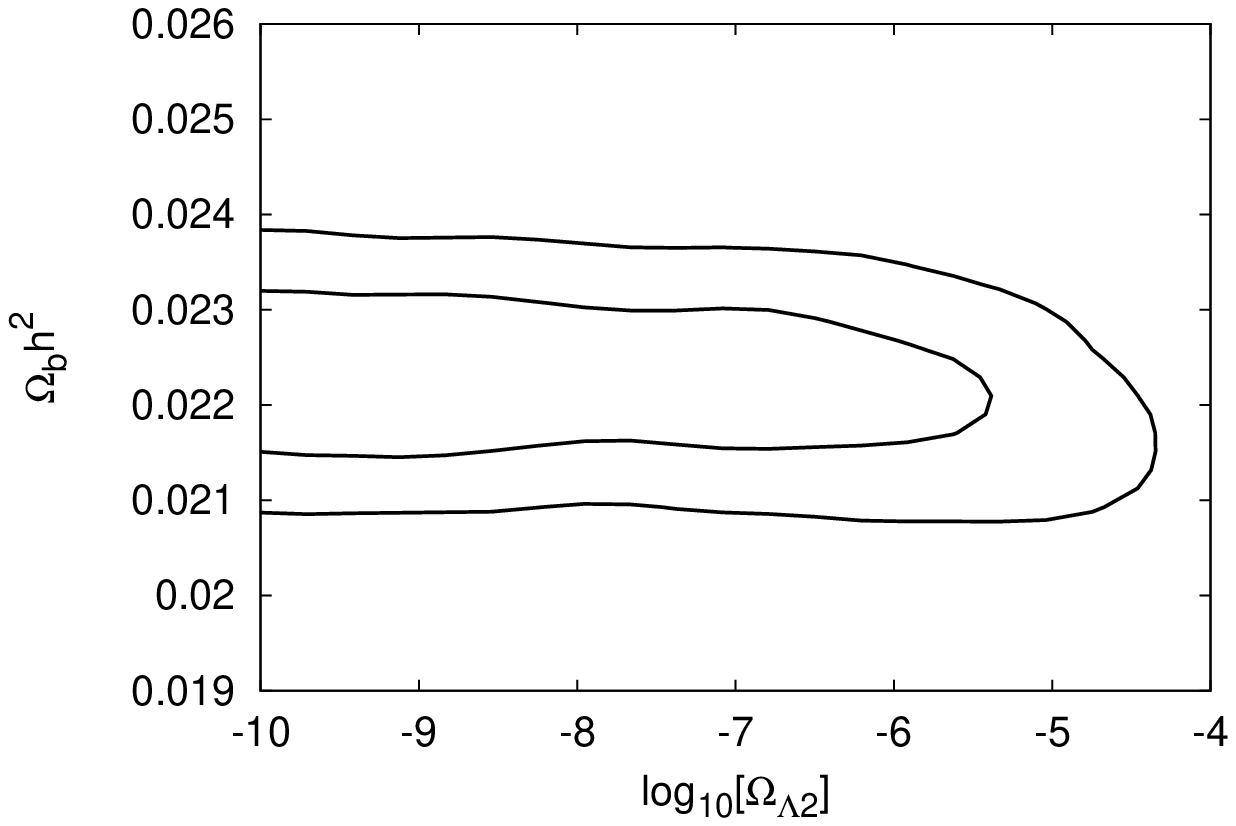}    
 \includegraphics[width=1.0\linewidth]{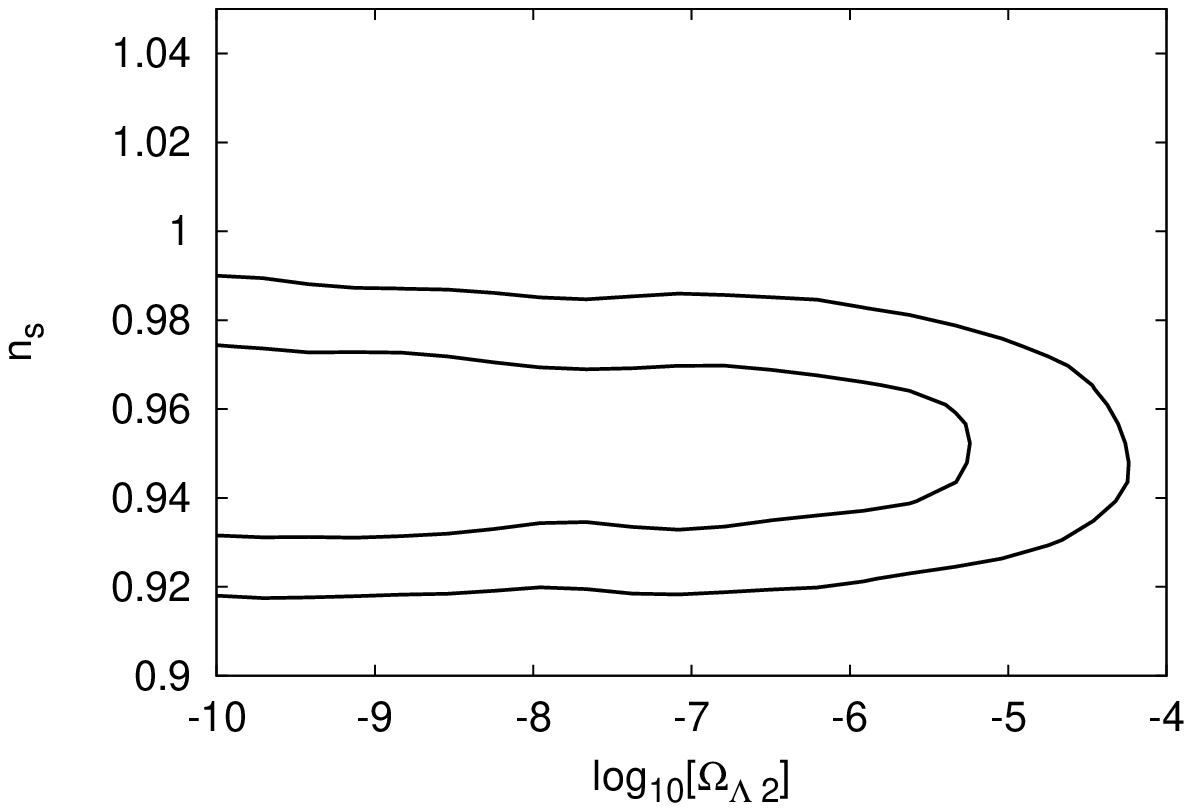}    
 \includegraphics[width=1.0\linewidth]{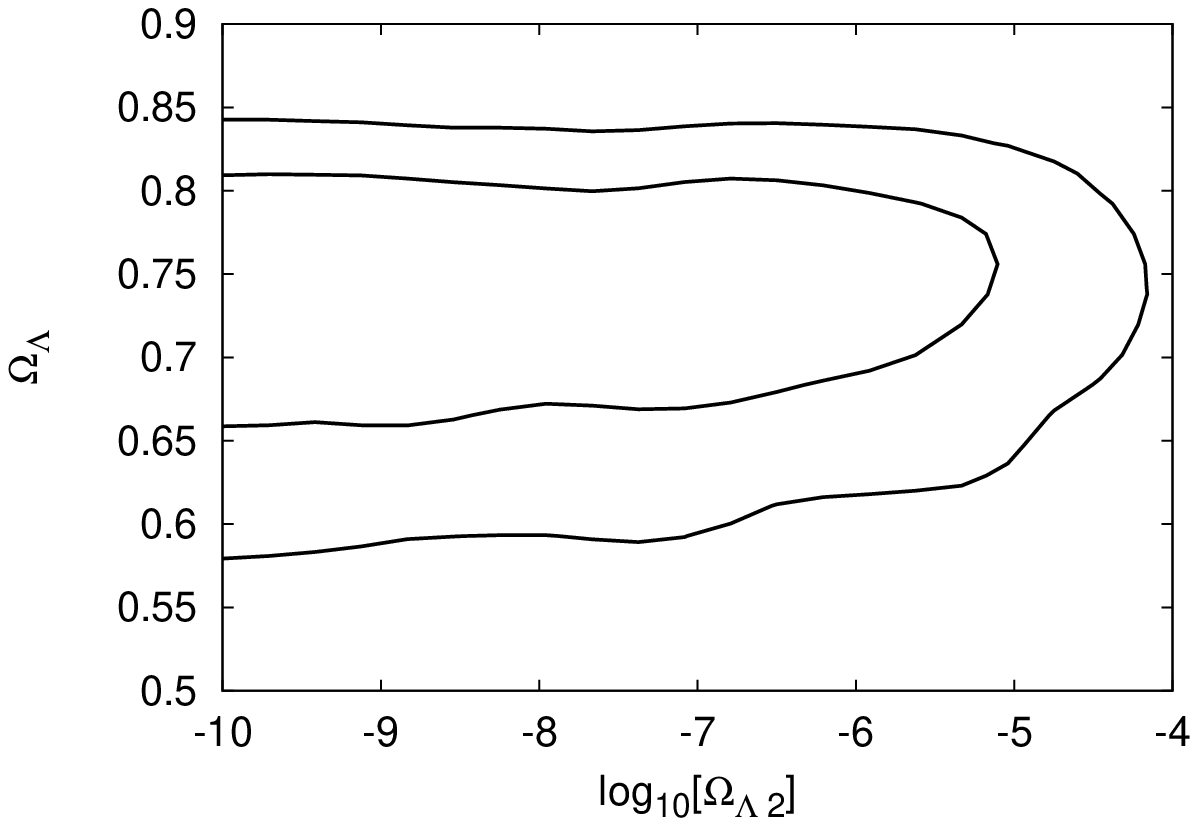}    
\end{minipage}    
\begin{minipage}{0.49\linewidth}    
 \includegraphics[width=1.0\linewidth]{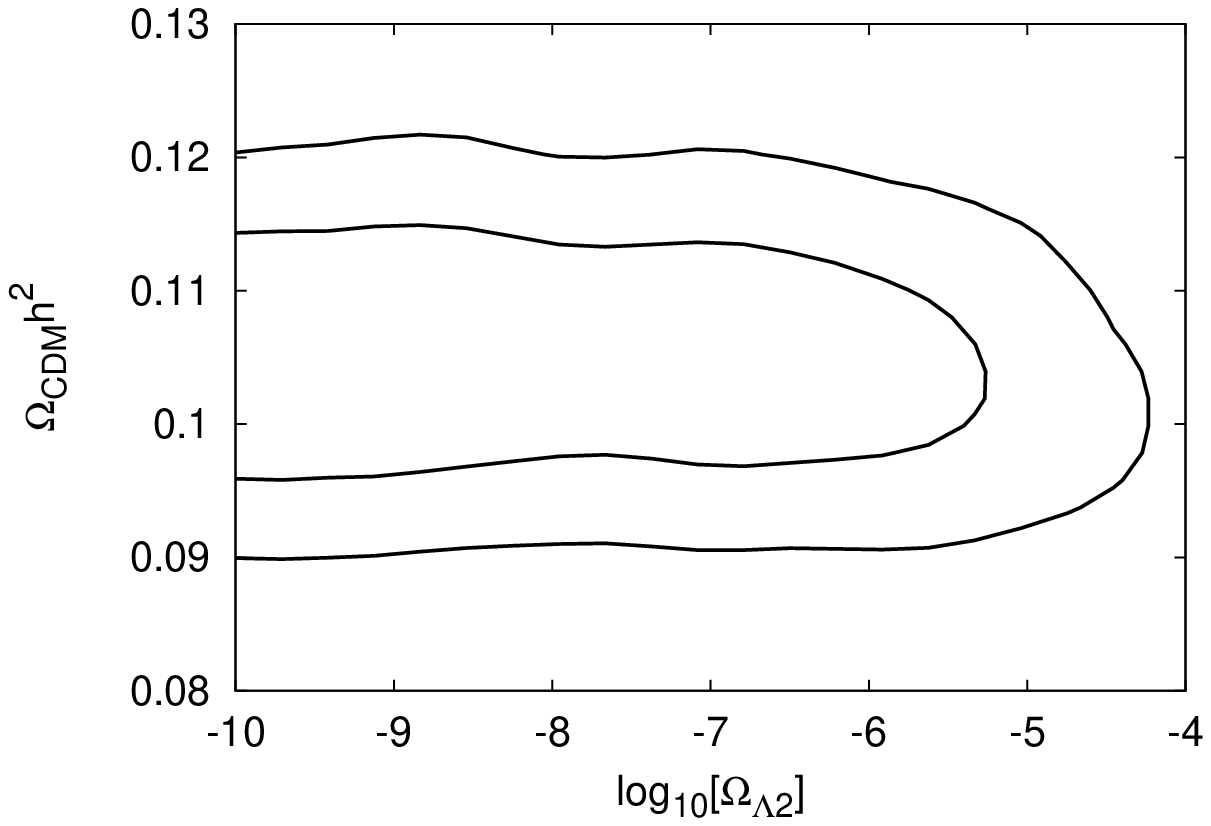}    
 \includegraphics[width=1.0\linewidth]{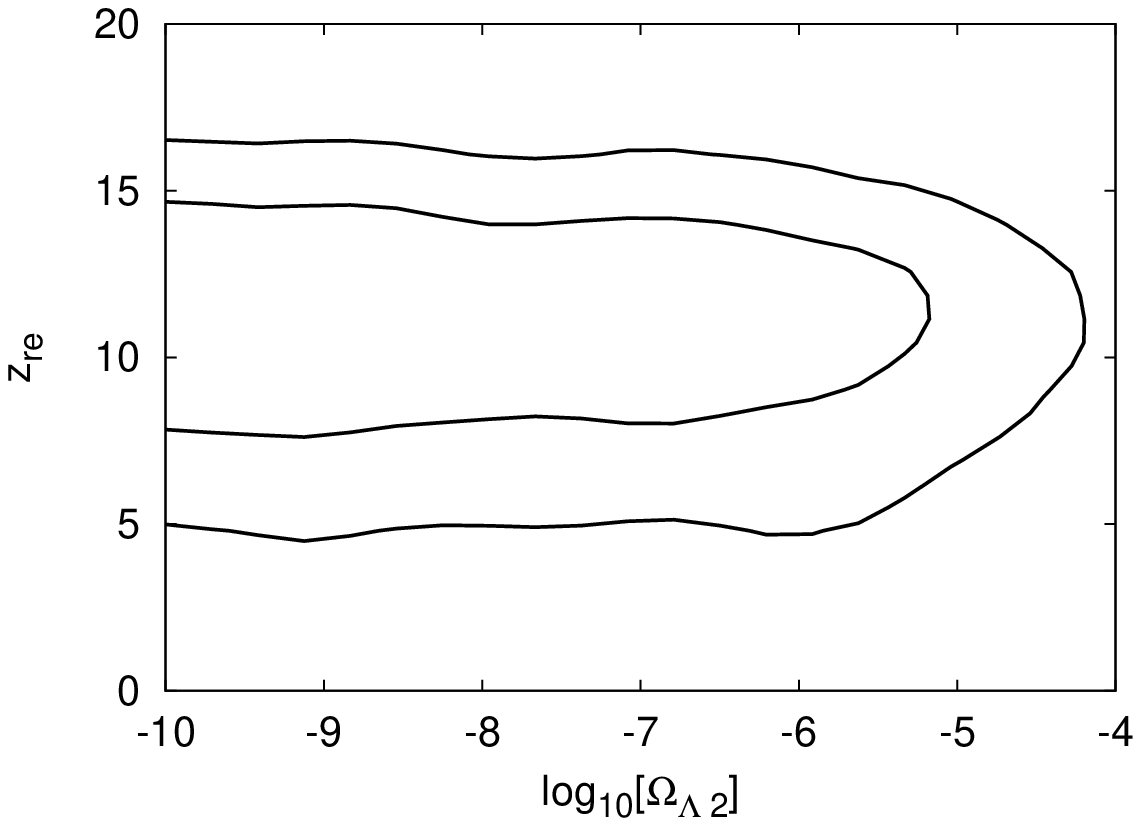}    
 \includegraphics[width=1.0\linewidth]{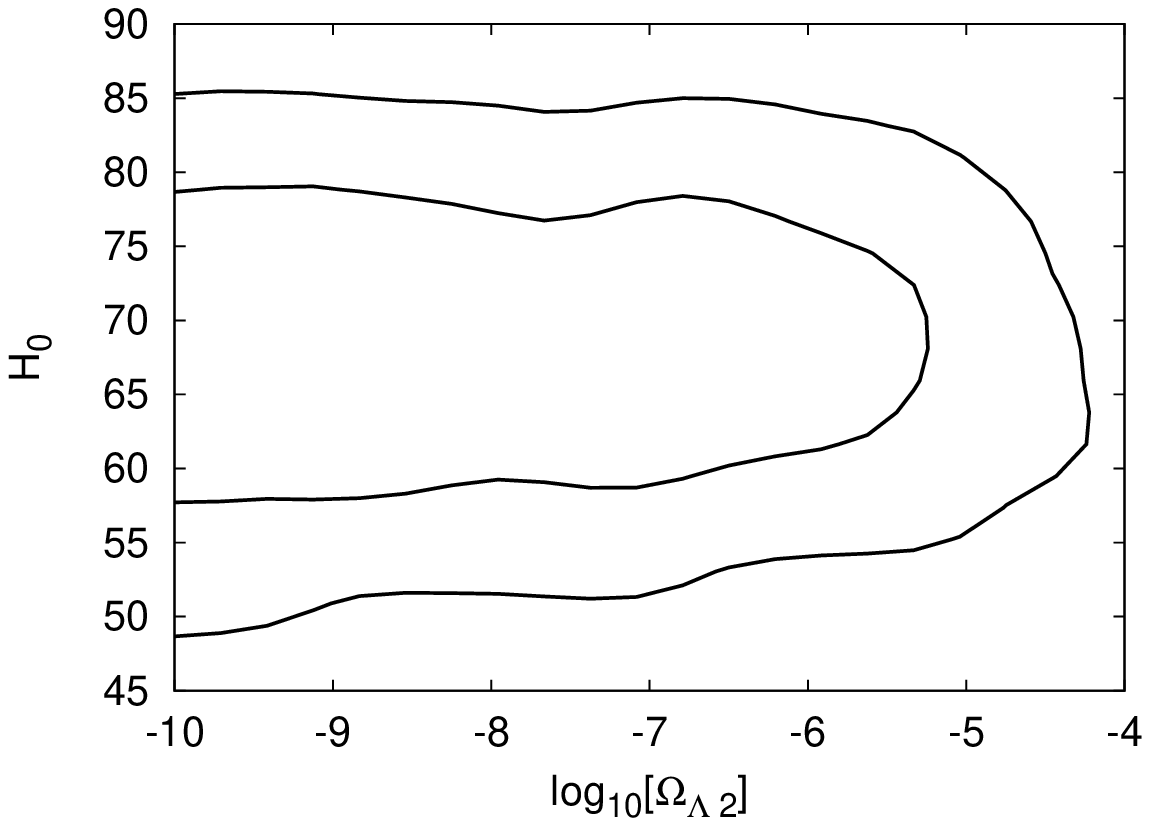}    
\end{minipage}    
\caption{\label{fig:contour3rd1} 
Constraints on $\log{\Omega_{\Lambda2}}$ against $\Omega_{b}h^2$ (left top panel),
 $\Omega_{CDM}h^2$ (right top panel),     
 $n_s$ (left middle panel),
 $z_{re}$ (right middle panel),    
 $\Omega^{}_{\Lambda1}$ (left bottom panel),    
 and $H^{}_0$ (right bottom panel)    
from WMAP three year results.    
Two curves in each panel correspond to the 68.3\% and 95.4\% confidence levels, respectively.    
}    
\end{figure*}

We  constrain the relation between $\Lambda_2$ ($m$) and other parameters from     
the recent CMB observations of WMAP \cite{Hinshaw06,Page06}    
, BOOMERanG \cite{Jones2005}    
, CBI \cite{Readhead2004}     
, and Acber \cite{Kuo2002}.    
Our results are shown in Figs. 
\ref{fig:cntr_dlcdm} and \ref{fig:contour3rd1}.    
Figure \ref{fig:cntr_dlcdm} shows the constraints on the
$m-\Omega^{}_{\Lambda2}$ plane and
our constraint is severer than that from the observed radiation
temperature in Fig.\ref{fig:m_oml2_temp}.
In our analysis, we obtain upper limits of $m$ and $\Omega^{}_{\Lambda2}$     
such as $m \le 4.2$ and     
$\Omega^{}_{\Lambda2} \le 1.7\times10^{-4}$ at the 95.4 \% confidence
levels, respectively. Therefore, we cannot find the clear evidence of a decaying-$\Lambda$.    
Figure \ref{fig:contour3rd1} shows the contours between $\Omega^{}_{\Lambda2}$    
and other cosmological parameters     
$(\Omega^{}_bh^2_{},\Omega^{}_{CDM}h^2_{}, n^{}_s, z^{}_{re} H^{}_0)$    
at the $68.3\%$ and $95.4\%$ confidence levels.    
Parameters in the D$\Lambda$CDM model    
have no degeneracy with other cosmological parameters;
the parameters are independent on other parameters.    
Table \ref{tab:parameters} shows the comparison of cosmological
parameters between the S$\Lambda$CDM and D$\Lambda$CDM model obtained
from our MCMC analysis. 
As expected from Figs. \ref{fig:cntr_dlcdm} and \ref{fig:contour3rd1}
differences in the cosmological parameters are as small as some percents.
However, the value of $\Omega^{}_\Lambda$ differs around $10$ percents,
which should be further constrained by future observations.


\section{\label{sec:rad} Summary and Discussion}    
    
 We have investigated the possible difference in the thermal evolution of the    
universe with a decaying-$\Lambda$ term as a function of the cosmic
scale factor that reduces the photon energy density.    
Although the energy density of a $\Lambda$-term is increasing at the    
early era,     
the effects of the decaying $\Lambda$ on the cosmic expansion rate can
be ignored at the $\Lambda$ dominate epoch. On the other hand, a decaying-$\Lambda$ term
has been insisted to alter the evolution of the photon temperature \cite{Kimura,Kamikawa,Puy}.     
Depending on parameters in the D$\Lambda$CDM model, the photon energy
density could be lower or higher compared with that in the S$\Lambda$CDM    
model at  $z>0$.     
However, the second case should not occur, because    
the photon temperature becomes negative at some epoch of $z<0$.    
We estimate the epoch of the last-scattering surface by calculating the
visibility function, and find that $z_{dec} = 2040$ for $m=1.2$ and
$\Omega^{}_{\Lambda2}=10^{-4}_{}$ in the D$\Lambda$CDM model, which indicate
that the decoupling occurs earlier by    
$\Delta z_{dec} = 950$ compared to the case of the S$\Lambda$CDM. 
    
We examine qualitatively the effects of a decaying-$\Lambda$ term on CMB
angular power spectrum.    
We obtain the modified Boltzmann equation of photons in the D$\Lambda$CDM model    
based on the cosmological perturbation theory and calculate the CMB
angular power spectrum. We find that a decaying-$\Lambda$ could alter
the CMB angular power spectrum 
significantly due to the following reasons:     
 large baryon energy density relative to the photon density causes to    
boost up the first and third peaks; the  early photon decoupling shifts
CMB spectrum to higher multipoles.
    
Finally, using the Markov-Chain Monte Carlo analysis,     
we can put constraint on $m$, $\Omega^{}_{\Lambda2}$, and cosmological 
parameters. 
We obtain the upper limits of parameters in    
D$\Lambda$CDM: $m<4.2$ and $\Omega^{}_{\Lambda2}< 1.7\times10^{-4}$
\footnote{Recently, WMAP five year results have been released \cite{Hinshaw2008},
where the third peak of the spectrum is clearly determined.
However differences in the first peak of WMAP five year compared to the
three year results are within 1$\sigma$ level, and 
those in the third peak are still more than 1$\sigma$ level.
Therefore if we reanalyze with the new WMAP data, our results
would not be affected within the present observational error of WMAP
five year}.
In our analysis, the upper limit of $m$ is close to 4.
A decaying-$\Lambda$ might affect primordial nucleosynthesis 
because the $\Lambda$-term evolves as radiation (photon, neutrino, and
electron-positron) at the early universe.
However, if $m$ is large, $\Lambda_2$ tends to be small as shown in Fig.\ref{fig:cntr_dlcdm}.
As the result, $\Lambda$-term becomes smaller than radiation components. 
In fact, at $T=10^9_{}$ K,
the ratio of the $\Lambda$-term and radiation density is $10^{-5}_{}$
for $m=4.2$ and $\Omega^{}_{\Lambda2}=10^{-10}_{}$ (upper limits from WMAP results).
Therefore, we can say that effects of the early universe such as
nucleosynthesis is small.   
Therefore, the effects of the $\Lambda$-term on the physical processes
in the early universe, such as on the nuclosynthesis, is negligible \cite{Ichiki2002}.

Interestingly, there is no degeneracy between the two parameters in    
D$\Lambda$CDM and other cosmological parameters.
From these constraints, the contribution of a decaying-$\Lambda$ term to the cosmic    
thermal evolution should be extremely small, 
since the best-fit values of $m, \Omega^{}_{\Lambda2}$ are nearly zero.
    
We assume a cosmological term as a function of scale    
factor for simplicity Even if we parameterize reasonably the evolution
of the cosmological term or the equation of state of dark energy,     
our results would not change qualitatively.    
    
On the other hand,     
we find that the reionization occurs at $z_{re}=11$ in the D$\Lambda$CDM model,     
which suggests that a first object could be formed at around this epoch.    
We should note that we assume that the reionization history can be
described by step-function as discussed in Ref. \cite{WMAP2006}.
The next CMB satellite, Plank, is expected to determine detailed reionization history. 
Then a variable $\Lambda$-term model such as D$\Lambda$CDM should be constrained further.

\begin{table} 
\caption{Comparison of cosmological parameters between the S$\Lambda$CDM and
 D$\Lambda$CDM models obtained from Markov-Chain Monte Carlo analysis.} 
\label{tab:parameters}     
\begin{ruledtabular}    
\begin{tabular}[t]{ccc}    
 & D$\Lambda$CDM &  S$\Lambda$CDM \\    
\hline    
$\Omega^{}_{\Lambda2}$  &    
$< 1.7\times10^{-4}$ &  \\    
$m$ &    
$<4.2$ & \\    
\hline    
$\Omega_b h^2$ &    
$0.0221^{+0.0019}_{-0.0028}$ & $0.0223\pm 0.0007$ \\    
$\Omega_{CDM} h^2$ &    
$0.103^{+0.021}_{-0.018}$ & $0.1037\pm 0.0081$ \\    
$\log_{10}{(10^{10}A_s)}$ & $2.991^{+0.019}_{-0.015}$ & $3.156\pm0.056$ \\    
$z_{re}$  & $10.6^{+6.2}_{-8.0}$ & $10.9^{+2.6}_{-2.7}$ \\    
$n_s$ & $0.945^{+0.050}_{-0.031}$ & $0.951\pm0.016$ \\    
$H^{}_0$ & $70.7\pm19.7$ & $71\pm3$ \\
\hline
$\Omega_\Lambda$ & $0.757^{+0.083}_{-0.206}$ & $0.763\pm0.034$ \\    
$\Omega_m$ & $0.249^{+0.261}_{-0.099}$ & $0.233^{+0.033}_{-0.034}$ \\    
\end{tabular}    
\end{ruledtabular}    
\end{table}    
    
\begin{acknowledgments}    
We would like to express our appreciation to K. Arai and T. Teranishi for useful discussion. 
This work has been supported in part by a Grant-in-Aid for Scientific
Research (18540279) of the Ministry of Education, Science and Culture in
Japan.
Data analysis was in part carried out on a general common user
computer system at the Astronomical Data Analysis Center of the National
Astronomical Observatory of Japan.
\end{acknowledgments}


\end{document}